\documentclass{optica-article}

\journal{oe}

\usepackage{url}
\usepackage{float}
\usepackage{lineno}
\usepackage{blindtext}
\usepackage{graphicx}
\usepackage{caption}
\usepackage{subcaption}
\usepackage{overpic}
\usepackage{makecell}
\restylefloat{table}
\usepackage{cleveref}
\usepackage{hyperref}
\hypersetup{hidelinks,colorlinks={false},linkcolor=black,citecolor={black}}
\usepackage{amsmath}
\usepackage{newtxtext,newtxmath}
\usepackage[version=4]{mhchem}

\begin{document}

\title{Optical convolutional neural network with atomic nonlinearity}

\author{Mingwei Yang,\authormark{1,2*} Elizabeth Robertson,\authormark{1,2
} Luisa Esguerra,\authormark{1,2} Kurt Busch,\authormark{3,4} and Janik Wolters\authormark{1,2}}

\address{\authormark{1}Deutsches Zentrum für Luft- und Raumfahrt, Institute of Optical Sensor Systems, Rutherfordstraße 2, 12489 Berlin, Germany.\\
\authormark{2}Technische Universität Berlin, Straße des 17. Juni 135, 10623 Berlin, Germany.\\
\authormark{3}Humboldt-Universität zu Berlin, Institut für Physik, AG Theoretische Optik \& Photonik, 12489 Berlin, Germany.\\
\authormark{4}Max-Born-Institut, 12489 Berlin, Germany.}

\email{\authormark{*}mingwei.yang@dlr.de} %% email address is required; see note below about the corresponding author designation



\begin{abstract}
Due to their high degree of parallelism, fast processing speeds and low power consumption, analog optical functional elements offer interesting routes for realizing neuromorphic computer hardware. For instance, convolutional neural networks lend themselves to analog optical implementations by exploiting the Fourier-transform characteristics of suitable designed optical setups. However, the efficient implementation of optical nonlinearities for such neural networks still represents challenges. In this work, we report on the realization and characterization of a three-layer optical convolutional neural network where the linear part is based on a 4f-imaging system and the optical nonlinearity is realized via the absorption profile of a cesium atomic vapor cell. This system classifies the handwritten digital dataset MNIST with $83.96\%$ accuracy, which agrees well with corresponding simulations. Our results thus demonstrate the viability of utilizing atomic nonlinearities in neural network architectures with low power consumption.
\end{abstract}

%%%%%%%%%%%%%%%%%%%%%%%%%%  body  %%%%%%%%%%%%%%%%%%%%%%%%%%
\section{Introduction}
In recent years, convolutional neural networks (CNNs) have established themselves as a key method in computer vision tasks, with applications that range from fundamental studies in condensed-matter physics \cite{carrasquilla2017machine} and particle physics \cite{radovic2018machine} all the way to autonomous driving \cite{FUJIYOSHI2019244} and cancer detection \cite{Yamashita2018}. With the rising popularity of CNNs, significant concerns regarding their energy consumption relative to simpler network architectures have emerged. Specifically, about $\thicksim80\%$ of the inference time required by CNNs is utilized for carrying out the convolution \cite{Li2016} so that energy-efficient computing paradigms for computing convolutions have hence become an active field of research. Due to their inherent parallelism, potential for GHz modulation speeds and low energy consumption (when using only passive elements), free-space-optics implementations have been identified as an attractive possibility for analog computations of convolutions \cite{caulfield2010future,caulfield1989optical}. In fact, within the broader context of artificial neural networks, linear-optics implementations have been demonstrated based on diffractive materials \cite{Lin1004}, spatial light modulators (SLMs) \cite{chang2018hybrid,Spall:20,zhou2021large,goorden2014superpixel,bueno2018reinforcement}, ring resonators \cite{tait2014broadcast}, arrays of Mach-Zender interferometers \cite{Shen2017,ZhouZhaoWeiLiDongZhang+2019+2257+2267} and wavelength-division multiplexing techniques \cite{feldmann2019all}. For further information, we would like to refer to recent reviews \cite{sui2020review,nano11071683,DeMarinis:19} and the convolutional layer design that has recently been demonstrated by Miscuglio et al. \cite{Miscuglio:20}. However, the most efficient optical implementation of the nonlinearities required by neural networks remains an open question. Recently, Zuo et al. have demonstrated an optical nonlinearity by utilizing electromagnetically induced transparency in a gas of ultra-cold \ce{^{87}Rb} atoms \cite{Zuo:19}. Ryou et al. have avoided the overhead associated with ultra-cold atoms by realizing a nonlinearity through saturable absorption in a thermal vapor of Rb atoms \cite{Ryou:21}. Other mechanisms for realizing optical nonlinearities for neuromorphic applications include phase-change materials (PCMs) \cite{feldmann2019all} and the Kerr effect combined with two-photon absorption \cite{mesaritakis2015all}.

In this work, we introduce a nonlinearity in the form of a saturated absorption profile of cesium atomic vapor into an optical convolution setup based on free-space SLMs. As we shall demonstrate below, this system classifies the handwritten digit dataset MNIST \cite{lecun1998mnist} with $83.96\%$ accuracy. As an essential part of the experimental development of multi-layer optical neural networks, the optical nonlinearity is proven to be effectively provided by the cesium atomic vapor. With only one stable isotope and a vapor pressure of $\thicksim 2 \cdot 10^{-6}$\:Torr at room temperature, the Cs D lines show excellent absorption properties, enabling pronounced nonlinearities without isotopic purification or power-consuming cell heating. This simplicity in the use of Cs cell is particularly important as we have developed the system with an eye toward space applications. In fact, we expect a potential benefit of optical neural networks with nonlinearity for data processing on board of satellites. The standard procedure for processing complex sensor data is the use of artificial neural networks, which is served on the ground by specialized digital hardware such as graphics cards, tensor flow processors, etc. The availability of these options is limited for data processing in orbit due to the extreme requirements for energy consumption, thermal management and radiation hardness. However, data transfer and processing on the ground can also be challenging due to the immense amount of data. Therefore, high-performance computers under space conditions would be desirable and optical computers have a high potential to fill the gap, enabling energy-efficient machine learning in orbit.

\section{Methods}

Here we demonstrate an optical convolutional neural network (OCNN) in which both linear operations and the nonlinearity are realized optically. We implement the OCNN with one input layer, one optical convolutional layer, one fully connected layer followed by one output layer. An optical nonlinearity is applied after the convolutional layer [Fig. 1]. The convolution of the input and the kernel of the OCNN is performed optically by pointwise multiplications in the Fourier plane, based on a 4f-imaging system and the convolution theorem. A CMOS camera
\begin{figure}
    \centering
    \includegraphics[width=\textwidth]{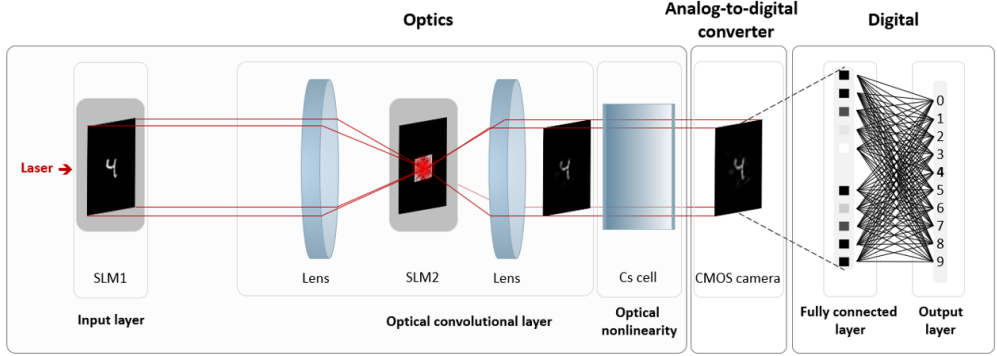}
    \caption{Layout of the optical convolutional network architecture used in this work.}
    %\label{fig:my_label}
\end{figure}
acts as an analog-to-digital converter. A digital fully connected layer is implemented in the computer to connect the nonlinear activated feature maps and the output layer. We simulate the system based on the experimental setup. In the simulation, the kernel of the OCNN and the digital fully connected layer are pretrained and applied to the experiment. The performance of the simulation and the experimental results are compared.

We train and test the system with handwritten digits from the MNIST dataset. The input data is encoded as a two-dimensional intensity profile by an amplitude-only SLM, which is based on a digital micromirror device (DMD). This electro-optic conversion is implemented by controlling the individual micromirrors of the DMD. These mirrors can rapidly tilt between an on-and-off position and selectively reflect the incident light to the optical path of the OCNN. One micromirror and its corresponding memory unit constitute one pixel of the DMD. The properties of the large-scale display ($1920 \times 1080$ pixels) and the high update rate up to $10 ^ {4}$\:Hz for binary images of the micromirror array allow in principle complex multi-channel data processing at high speed.

 The conventional convolution process is formed of the pixel-wise multiplication and summation of a subsection of an image with a kernel. The kernel is scanned across the whole image, where this multiplication and summation repeats, resulting in an image convolved with a kernel. By contrast, the OCNN performs the convolution process by employing the convolution theorem $ f(x) \ast h(x) = \mathcal{F}^{-1} \{ \mathcal{F}[f(x)] \cdot \mathcal{F}[h(x)]\} $ and the Fourier transform properties of lenses \cite{goodman2005introduction}. Specifically, the convolution ($ \ast $) of the input $f(x)$ and the kernel $h(x)$ is the inverse Fourier transform ($\mathcal{F}^{-1}$) of the pointwise product ($\cdot$) of their Fourier transforms ($\mathcal{F}[f(x)],\mathcal{F}[h(x)] $). By constructing a 4f-system with two SLMs and two lenses, the Fourier transform of the input, the dot product in Fourier space and the inverse Fourier transform of the dot product can be performed optically and passively. Thus, the convolved images can be observed in the front focal plane of the second lens.
 
\begin{figure}[t]
    \centering
    \includegraphics[width=0.8\textwidth]{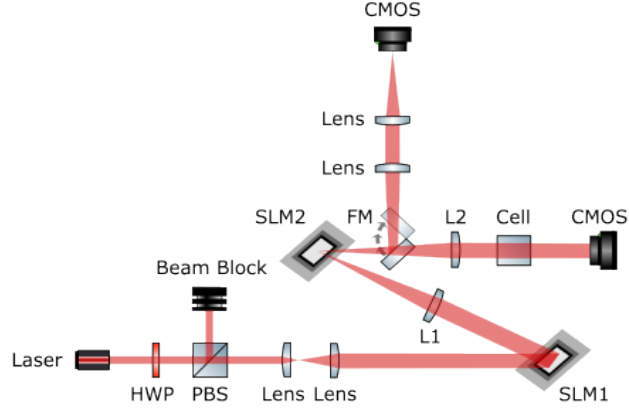}
    \caption[Sketch of the experimental setup.]
    {Sketch of the experimental setup. M: mirror, HWP: half-wave plate,
PBS: polarizing beam splitter, L: lens, FM: flip mirror. The intensity of light is controlled by the combination of the half-wave plate and the polarizing beam splitter. A telescope is composed of a 50\:mm and a 100\:mm focal length lens, expanding the beam to 8\:mm diameter. 
}
    %\label{fig:my_label}
\end{figure} 

In the experiment, we place the cesium vapor cell after the convolutional layer, so that the convolved pattern is nonlinearly absorbed, thereby introducing an optical nonlinearity into the system. The CMOS camera is positioned in the front focal plane of the second lens. The images are captured by the camera and saved on a computer. Figure 2 depicts the sketch of the experimental setup. The light source is a distributed feedback diode laser assembly with a Doppler-free spectroscopy setup. The wavelength is fine-tuned and actively stabilized to the cesium D1 line transition ($6^2S_{1/2}$ $\rightarrow$  $6^2P_{1/2}$) at about $894$\:nm. The laser is further coupled into and emerges from an optical fiber. A half-wave plate and a polarizing beam splitter are used to adjust the light intensity. Thereafter, two lenses act as a telescope to enlarge the beam diameter to approximately 8\:mm. This large collimated beam is sufficient to cover the pattern surface of SLM1 (DLP6500, Texas Instruments, pixel size $7.6\:{\rm \mu m}\times 7.6\:{\rm \mu m}$). A computer connects and loads the patterns to SLM1, therefore, the incoming beams are specifically modulated to the image shapes of the binarized MNIST dataset and reflected to the optical path. For the used SLM model, a theoretical maximum pattern rate of $100 \:\rm kHz$ can be reached with binary data. A lens ($ f= 250\:\rm mm$, L1) is located at a distance of 250\:mm from SLM1. The second identical SLM is then placed in its back focal plane, multiplying the corresponding Fourier-transformed input with the displayed Fourier-transformed kernel. The amplitude modulated dot products of the Fourier patterns are then inverse Fourier transformed by a second lens ($ f= 250\:\rm mm$, L2), and subsequently captured by the CMOS camera (CS165MU- 1.6 MP Monochrome CMOS Camera, Thorlabs, pixel size $3.45\:{\rm \mu m\times 3.45\:\rm \mu m}$) after being nonlinearly activated by the Cs vapor cell. Consequently, the images captured by the camera are the nonlinearly activated convolutions of the input images and the kernel. The computer collects all such processed images and converts them into arrays of gray scale values. Each array is connected to a trained fully connected layer, which outputs the prediction of the neural network via matrix multiplication. In addition, a flip mirror can be used to reflect the beam to a second CMOS camera for the purpose of imaging the Fourier pattern displayed on SLM2.

As discussed above, much attention has been drawn to optical nonlinearities. Atomic vapor cells are competitive options due to their non-vacuum, room-temperature experimental environment. Moreover, as a result of the passive pass of laser light, there is no need for additional energy. The relation between the output intensity $I$ and the input intensity $I_0$ is \cite{foot2004atomic}
 \begin{equation}
 \begin{split}
     %{I} = {I_0}\,{\rm exp}\bigg(\frac{-\mathrm{\it{{OD}_0}}}{1+{I_0}/{I_s}}\bigg),
     {I} = {I_0}\,{\rm exp}\bigg(\frac{-\mathrm{{{OD}_0}}}{1+{I_0}/{I_s}}\bigg),
     \end{split}
 \end{equation}
where $I$, $I_0$, $I_s$ represent the output intensity, input intensity, and saturation intensity, respectively. The optical depth at $I_0 = 0$ is given by $\mathrm{{OD_0}} = N\sigma z$, where $N$ is the number of cesium atoms per unit volume, $\sigma$ denotes the corresponding cross-section, and $z$ is the cell length. When the input intensity $I_0$ reaches the saturation intensity $I_s$, the absorption is reduced by $50\%$. The input-output relation thus follows a nonlinear shape. In the experiment, the light intensities are converted to 8-bit arrays of integer values by the camera. To ensure that the corresponding simulation is consistent with the experiment, we experimentally determine the input-output relation. To this end, we display a square pattern on SLM1 representing the average number of bright pixels in the MNIST dataset. While varying the incident laser power, we measure the resulting pixel intensity detected by the CMOS camera with and without the Cs cell, respectively. The measured and fitted input-output relation of the nonlinearity is shown in Fig. 3.

\begin{figure*}
\centering
\includegraphics[width=0.7\linewidth]{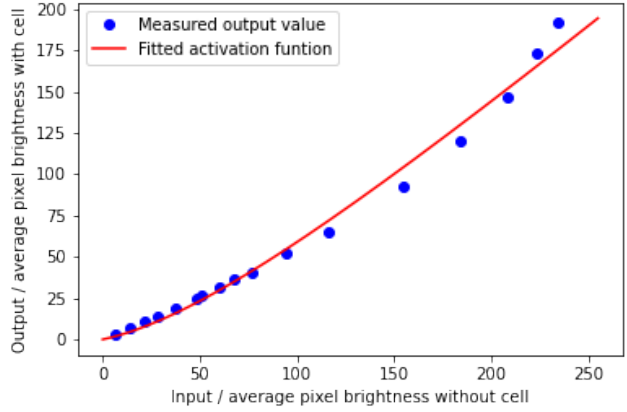}
\label{fig:nonlinearcam}
\captionsetup{skip={8pt}}  
\caption{Measured input-output average pixel brightness and the corresponding fit curve. Fitting the experimental data to Eq. (1) gives $\mathrm{\it{OD_0}} = 1.38, I_\text{s} = 62.31$ pixel brightness. The axes represent the average grayscale value of the picture without (input) and with (output) the vapor cell, respectively.}
\label{fig:Nonlinearity}%
\end{figure*} 

As a programming language, we have employed Python (version 3.8.5.) in combination with the Pytorch framework (version 1.7.1)\cite{pytorch} for training. This combination features an autograd function that nicely supports our custom nonlinear activation function during the backpropagation process. Contrary to a conventional CNN training process, different methods and additional procedures must be considered in the simulation in order to fit the optical setup. The OCNN model has a three-layer structure including the atomic nonlinearity. The original image size of the MNIST is $28\times28$, however, for experimental convenience, the input size is enlarged to $256 \times 256$. Moreover, according to the convolution theorem, the kernel size must be same as the input size, so the kernel size is also set to $256 \times 256$. The input and kernel are binarized for the purpose of effectively interacting with the cesium atoms. In the OCNN model, the convolution stems from the inverse Fourier transform of the multiplication of the Fourier-transformed input and the kernel. Thus, we have implemented a custom convolutional layer into the Pytorch machine-learning framework \cite{pytorch}. Due to the fact that we initialize the learnable kernel values directly in complex Fourier space \cite{pratt2017fcnn}, no Fourier transform is required for the kernel during the training process. The real-part is composed of random positive values. As amplitude-only SLMs do not deal with the phase information, all the imaginary values are set to zero. The first step is to perform a two-dimensional Fourier transform of the input images and multiply them with the kernel in Fourier space. Although the real-part values of the kernel are initialized to be positive, they may contain negative values during training. As a result, we apply a rectified linear unit (ReLU) function to the kernel to replace negative values with zero. Furthermore, we ensure that the pixel position of the Fourier pattern of the input light has a one-to-one match to the pixel position of the kernel structure on the SLM2. However, exact matching is experimentally difficult for the SLM hardware that has micron-scale pixels. Accordingly, the $256 \times 256$ kernel is max-pooled to $32 \times 32$, followed by a process of enlarging back to $256 \times 256$. Next, the element-wise multiplication is performed in Fourier space, and the following inverse Fourier transform gives the feature map of the OCNN in real space, the size maintains $256 \times 256$. It is worth mentioning that, in order for the optical patterns to be activated in a nonlinear manner, the light intensity is adjusted to that the input intensity of the vapor cell falls within the nonlinear region around the saturation intensity. The feature maps are max-pooled to $28 \times 28$ and flattened to one-dimensional arrays, then a fully connected layer with a size of $784 \times 10$ further processes the feature maps and outputs a $10 \times 1$ array representing the predicted digits. The model is trained for 5 epochs, using a gradient descent optimization algorithm with the Softmax classifier as loss function, Adam \cite{kingma2014adam} as the optimizer, and $0.001$ learning rate.

\section{Results} 

We have performed the training of our OCNN with the $60,000$ training images from MNIST and have determined the corresponding accuracy for $10,000$ test data. For the MNIST test set, the simulation of the OCNN achieve $ 92.6\% $ and $ 93.2\%$ accuracy with the cesium absorption function and the ReLU function as nonlinearity, respectively. The accuracy is relatively low for the MNIST dataset comparing to other SLM-based optical network systems \cite{zhou2021large, Miscuglio:20}, since the dataset is binarized in the OCNN and there is only one convolutional layer with a single kernel for the first demonstration and better reproducibility of the training. Without any nonlinearity in the entire network, the accuracy goes down to $90.8\%$. For comparison, a conventional feed forward model is trained ($5$ epochs, $0.001$ learning rate). This model which consists of one fully connected layer of size $784\times 10$ gives an accuracy of $80.5 \%$. 

The experimental classification accuracy for both cases, with and without the vapor cell are tested directly using the individually simulated kernel and fully connected layer. Figure 4(a) shows examples of simulated and experimentally captured feature maps. For the absence of the vapor cell, the OCNN setup achieves $70.72\%$ accuracy for test data. With the cesium vapor cell in the setup, the performance is improved to $71.84\%$. The confusion matrix indicates that the experimental results have particular errors in predicting the digits "1", "5", and "8" [Fig. 4(b)]. The comparable low accuracy is caused by experimental imperfections. In order to compensate for these imperfections, we train a mapping matrix using the full $10,000$ test data being processed by the optical system. This linear mapping is used to modify the fully connected layer, i.e. to retrain it to the experimental data. As shown in the confusion matrix, the trained mapping matrix significantly reduces mispredictions [Fig. 4(c)]. With the help of
\begin{figure}[h]
     \centering
     \begin{subfigure}[b]{0.85\textwidth}
         \centering
         \includegraphics[width=0.85\textwidth]{figure-4a.pdf}
         \caption{}
         %\label{fig:y equals x}
     \end{subfigure}
     %\vspace{10mm}

     \begin{subfigure}[b]{0.35\textwidth}
         \centering
         \includegraphics[width=0.98\textwidth]{figure-4b.pdf}
         \caption{}
         %\label{fig:three sin x}
     \end{subfigure}
     %\hfill
     \hspace{10mm}
     \begin{subfigure}[b]{0.35\textwidth}
         \centering
         \includegraphics[width=0.98\textwidth]{figure-4c.pdf}
         \caption{}
         %\label{fig:five over x}
     \end{subfigure}
        \caption{Visualization of feature maps and confusion matrices. (a) Examples of simulated (first row) and experimental (second row) feature maps. (b) Experimental confusion matrix of the MNIST classification. (c) Confusion matrix of the MNIST classification with the trained mapping matrix.}
        \label{fig:three graphs}

        \vspace{5mm}
        \hfill
        \begin{minipage}[b]{\textwidth}
        \centering
        \setlength{\tabcolsep}{2.3mm}
        \begin{tabular}{c|c|c|c}
        Non-linearity & Cs Vapor & ReLU & No nonlinearity\\ 
        \hline
        Simulation & $ 92.6\%$ & $ 93.2\%$ & $ 90.8\%$ \\ 
        \hline
        OCNN (No mapping matrix) & $ 71.84\%$ & $ 70.72\%$ & $ 45.59\%$\\ 
        %\makecell{OCNN \\(No FC training)} & $ 71.84\%$ & $ 70.72\%$ \\ 
        \hline
        OCNN (With mapping matrix) & $ 83.96\%$ & $ 91.9\%$ & $ 89.3\%$\\
        %\makecell{OCNN \\(With FC training)} &$ 83.9\%$ & $ 91.9\%$ \\
        \end{tabular}
        
         %setlength{abovecaptionskip}{1.cm}
         %vspace{50mm}
         \captionsetup[table]{skip=10pt}
         %caption{Summary of OCNN Prediction Accuracy}
         \captionof{table}{Summary of OCNN prediction accuracy. Each kernel is individually trained for each case. No nonlinearity means that no additional nonlinearity is added beyond the inherent nonlinearity of the camera.}
        \end{minipage}

\end{figure}
\noindent this training, the accuracy increases to $ 83.96\%$ and $ 91.9\%$ as a result of with and without the vapor cell, respectively. For comparison, the accuracy of the OCNN without any additional nonlinearity gives an accuracy of $89.3\%$, indicating the effectiveness of the nonlinearity introduced by the camera. In Table 1. we have summarized the results on the accuracy. After retraining the mapping matrix, the experimental classification performance is in good agreement with the simulation. We attribute the slightly lower performance of the optical system to the reasons: 1) As the dynamic range and analog-to-digital conversion accuracy of the camera are limited, the distribution of pixel values does not accurately match the theoretical values. 2) The Fourier-transformed input images and the kernel image displayed on SLM2 do not overlap perfectly. 3) Distortions and aberrations caused by optical imperfections of the optical components also affect the experimental results.

\section{Conclusion}

In summary, we have demonstrated an optical convolutional neural network with atomic nonlinearity. The linear part of the system consists of a 4f-system made from SLMs and lenses, the optical nonlinearity is realized by a cesium vapor cell. The prediction accuracy of the OCNN setup for recognizing the MNIST handwritten digit dataset reaches $83.96\%$, which is in reasonable agreement with the simulation ($ 92.6\% $). The effectiveness of the atomic vapor cell suggests that it has great potential to provide optical nonlinear activation functions for neural networks with other topologies. Moreover, it is an attractive candidate for use in multilayer optical neural networks with several SLMs and vapor cells in series, as the light beam passively acquires nonlinear output characteristics without requiring a further source of energy. In addition, the operation complexity is also relatively low and for further integrations, the atomic vapor cell can also be replaced by semiconductor saturable absorbers (SESAM) \cite{dejonckheere2014all}.

For the specific task of classifying MNIST, the nonlinearity has only limited impact. However, our results show the great potential of OCNN with nonlinearity for machine learning problems in general. The performance can theoretically reach $10^6\times10^6\times10^4=10\times10^{15}$ operations per second (OPS) with image and kernel size of $10^6$ each and binary resolution allowing for a maximum pattern rate of $10^4$ Hz. Then, bottlenecks will include the resolution and update rate of hardware, the transmission rate of interfaces, etc. As a comparison, Google's latest release of the state-of-the-art TPU v4 chip, provides up to $\thicksim 275 \cdot 10^{12}$ OPS. Regarding power consumption of the OCNN, the laser and camera together require only about $2$\:W, while the two SLMs need on the order of $10$\:W. The used computer requires about $50$\:W. However, the atomic nonlinearity enables the possibilities for realizing multilayer ONNs with all-optical training \cite{guo2021backpropagation}, so the computer might be omitted in the future. Moreover, the SLMs mainly consume power for switching. Thus, the numerous quasi-static SLMs that provide connectivity in such a multilayer ONN are expected to have negligible impact on the power consumption being dominated by the input SLM. Hence, we expect a multilayer optical neural network to drastically increase the computational performance without increasing the energy consumption significantly, thereby rendering optical neural networks with atomic nonlinearity an attractive alternative to conventional hardware.

\begin{backmatter}
\bmsection{Funding}
This work was funded by the Deutsche Forschungsgemeinschaft (DFG, German Research Foundation) – Project number 445183921. E.R. acknowledges funding through the Helmholtz Einstein International Berlin Research School in Data Science (HEIBRiDS).

\bmsection{Acknowledgments}
We thank G. Gallego for the discussions about the optical convolutional neural network architecture.

\bmsection{Disclosures}
The authors declare no conflicts of interest. 

\bmsection{Data availability} Data underlying the results presented in this paper are not publicly available at this time but may be obtained from the authors upon reasonable request.

%\bmsection{Supplemental document}

\end{backmatter}

%%%%%%%%%%%%%%%%%%%%%%% References %%%%%%%%%%%%%%%%%%%%%%%%%

%Add references with BibTeX or manually.
%\cite{Zhang:14,OSA,FORSTER2007,Dean2006,testthesis,Yelin:03,Masajada:13,codeexample}

%%%%%%%%%% If using BibTeX:
\bibliography{lib}

%%%%%%%%%% If preparing manually:
% \newcommand{\enquote}[1]{``#1''}

% \bibitem{Zhang:14}
% Y.~Zhang, S.~Qiao, L.~Sun, Q.~W. Shi, W.~Huang, L.~Li, and Z.~Yang,
%   \enquote{Photoinduced active terahertz metamaterials with nanostructured
%   vanadium dioxide film deposited by sol-gel method,}
%   {\protect\JournalTitle{Optics Express}} \textbf{22}, 11070--11078             (2014).

% {Optical Society}, \enquote{{OSA Publishing},}
%   \url{http://www.osapublishing.org}.

% \bibitem{FORSTER2007}
% P.~Forster, V.~Ramaswamy, P.~Artaxo, T.~Bernsten, R.~Betts, D.~Fahey,
%   J.~Haywood, J.~Lean, D.~Lowe, G.~Myhre, J.~Nganga, R.~Prinn, G.~Raga,
%   M.~Schulz, and R.~V. Dorland, \enquote{Changes in atmospheric consituents and
%   in radiative forcing,} in \enquote{Climate Change 2007: The Physical Science
%   Basis. Contribution of Working Group 1 to the Fourth assesment report of
%   Intergovernmental Panel on Climate Change,}  S.~Solomon, D.~Qin, M.~Manning,
%   Z.~Chen, M.~Marquis, K.~B. Averyt, M.~Tignor, and H.~L. Miler, eds.
%   (Cambridge University Press, 2007).

% \end{thebibliography}

\end{document}